\documentstyle[11pt,fleqn]{article}
\date{}
\def\picture #1 by #2 (#3){
  \vbox to #2{
    \hrule width #1 height 0pt depth 0pt
    \vfill
    \special{picture #3} % this is the low-level interface
    }
  }
\def\scaledpicture #1 by #2 (#3 scaled #4){{
  \dimen0=#1 \dimen1=#2
  \divide\dimen0 by 1000 \multiply\dimen0 by #4
  \divide\dimen1 by 1000 \multiply\dimen1 by #4
  \picture \dimen0 by \dimen1 (#3 scaled #4)}
  }
\begin{document}
\thispagestyle{empty}
\title{\bf Nontopological Global Field Dynamics}
\author{}
\maketitle
\begin{center}
{\bf Leandros Perivolaropoulos}\\
\vspace{.5cm}
Division of Theoretical Astrophysics\\
MS-51\\
Harvard-Smithsonian Center for Astrophysics\\
60 Garden St.\\
Cambridge, Mass. 02138 \\
{\it and}\\
Department of Physics \\
Brown University \\
Providence, R.I. 02912, U.S.A.
\end{center}

\begin{abstract}
We use arguments based on Derrick's theorem to
show that the property of collapse which is the key feature of
global texture appears in several field theory models with broken
global O(N) symmetry. Such models do not necessarily have nontrivial
third homotopy group of the vacuum manifold but may give rise to
collapsing global field configurations with properties similar to textures.
It is verified that
configurations with planar and cylindrical geometries do not
collapse. The existence of critical parameters for collapse of
spherically symmetric global field configurations is verified both
analytically and numerically.
\end{abstract}
\par
The texture model \cite{t89} for structure formation is simple and makes
several
encouraging predictions. In this model the primordial perturbations
that gave rise to structure formation are produced by the collapsing
knots of global field configurations.
\par
The case that has been well studied up to now is that of topological
$\pi_3$ textures \cite{t89}. These are spherically symmetric global field
configurations in models with broken O(4) symmetry.
They may become topologically nontrivial with appropriate boundary
conditions which compactify the physical space $R^3$ to the three sphere
$S^3$. Topological configurations could result by maps from the
compactified physical
space $S^3$ to the vacuum manifold $M={O(4)\over O(3)}=S^3$.
The homotopy group that leads to this nontriviality is $\pi_3(S_3)=Z$.
\par
In contrast to other topological defects (strings, walls, monopoles) it is not
the nontrivial
topology that makes the texture model interesting. It is the property  of field
collapse which leads to distinct nongaussian features in the produced
primordial perturbations. It is therefore important to investigate
this property in some detail. In particular, interesting questions to
address are the following: How generic is the spherical geometry of
collapse that has been observed in texture simulation so far? Does the
property of collapse persist in global fields with broken O(N) ($N\neq
4$) symmetry called {\it nontopological textures} \cite{t89} \cite{ts91}?
What are the critical parameters that determine whether
a global field configuration will collapse?
\par
This brief report is an attempt to shed some light to these questions. In
what follows we will consider global fields whose dynamics is
determined by the Lagrangian density:
\begin{equation}
{\cal L} = {{1 \over 2} {\partial_{\mu}\vec\Phi}\cdot{\partial^{\mu}\vec\Phi}}-
{{1\over 4} \lambda ({\vec\Phi}\cdot{\vec\Phi}-1)^2}
\end{equation}
where $\vec\Phi$ is a scalar field with O(N) symmetry and we have
normalized the scale of symmetry breaking to 1.
In studying its evolution, the condition
$\vec\Phi^{2}=1$ may be imposed on the scalar field.
This approximation
reduces the equations of motion for $\vec\Phi$ to those of the
nonlinear O(N)
model.
\begin{equation}
\eqalign{{\partial^{\mu}\partial_{\mu}\Phi^{i}}&= ({{\vec\Phi \cdot
{\partial^{\mu}\partial_{\mu}\vec\Phi}}})
\Phi^{i}\cr
{\vec\Phi}^{2}&=1\cr}
\end{equation}
with i=1,...,N. The above approximation is valid in regions of physical
space where the field gradient energy is smaller than the scale of
symmetry breaking. Therefore we can correctly follow the collapse of a
global field until the energy gradients become equal to the scale of
symmetry breaking. This is sufficient for our purpose which is to
determine the critical parameters for collapse.
\par
The method we use to determine the basic features of the
evolution of global fields is based on arguments used in the proof of Derrick's
theorem\cite{d64}.
 According to this theorem there are no static, finite energy  global
field configurations in space dimensions larger than two.
 Derrick's theorem however can not be applied
in a cosmological setup because by causality the field values will be
uncorrelated on horizon scales after a cosmological phase transition.
This lack of correlation will introduce diverging terms in the energy
coming from field angular gradients. Thus Derrick's theorem is
inapplicable. Here we modify the arguments of Derrick's theorem and apply them
to
field configurations with diverging energy. In particular we use a
rescaling of the global field coordinates and determine what type of
scaling is energetically favorable. Thus we distinguish between
configurations with expanding and collapsing volume of fixed energy (refered to
simply as expanding
and collapsing configurations). Our method attempts to compress a large amount
of information about
the evolution of a scalar field into a single rescaling parameter. It should
therefore be used
with care.In all cases presented here the analytical results have been
numerically
verified.
  \par
 In what follows we will discuss cases where the energy density has planar,
cylindrical or spherical symmetry. This simplification will allow us to obtain
both analytical and
numerical results thus testing both approaches.
\par
We start with the case of planar geometry.
The simplest nontrivial global field configuration with planar
geometry and the constraint ${\vec \Phi}^2=1$ appears in the
nonlinear O(2) model.
The field may be written as
$
{\vec \Phi}= (\sin \Psi_1(x,t),\cos\Psi_1(x,t))
$
with static energy
$
E_{st}={1\over 2}\int
dx \hskip .1cm ({\vec {\bigtriangledown_{}}}{\vec\Phi})^2=
{1\over 2}\int dx\hskip .1cm
\Psi_1^{\prime 2}(x,t)
$
and equation of motion
${\ddot \Psi_1}-{\Psi_1^{\prime\prime}}=0$.
Assuming that $\Psi^\prime$ is well behaved
at large x ( reasonable assumption even in a cosmological setup)
$E_{st}$ will not be diverging. It is easy to see that a
rescaling\cite{d64} of x in $\Psi_1$ ($\Psi_1(x)\rightarrow \Psi_1(\alpha x)$)
leads to a rescaling of $E_{st}$: $E_{st}\rightarrow \alpha E_{st}$.
Therefore dynamics will favor decrease of $\alpha$ leading to
expansion rather than collapse of the initial configuration. In fact
this O(2) case is particularly simple since $\Psi_1$ is a massless
Goldstone boson in one dimension and therefore
$\Psi_1(x,t)=\Psi_1(kx\pm
\omega t)$ with $k^2=\omega^2$.
Thus we expect the evolution to be that of two traveling
waves evolving in opposite directions and thus decreasing $\alpha$.
%\begin{figure}[htb]
%\hspace{1in}
%\scaledpicture 7.5in by 5.0in (fig1 scaled 600)
%\caption{The evolution of $\Psi_1(x,t_i)=0.9\pi(1-e^{-x})$ based on
 %O(2) model dynamics.}
%\end{figure}

This is demonstrated in Fig. 1 where we show the evolution of
$\Psi_1(x)=0.9\pi(1-e^{-x})$ on the half1ine $x\in[0,\infty)$. The
algorithm used is a leapfrog with fixed boundary conditions at the
origin and ref1ective at the other boundary. The same code with minor
modifications was used throughout this work.

\par
Field configurations with cylindrical geometry may also be shown not
to collapse. The case when the field has no angular dependence
\hfil\break
(e.g. ${\vec \Phi} = (\sin
\Psi_2(\rho,t)\sin\Psi_1(\rho,t),\sin\Psi_2(\rho,t)
\cos\Psi_1(\rho,t),\cos\Psi_2(\rho,t)$)
is similar to the planar case and may be shown not to lead to
collapse. We will here demonstrate the more interesting cosmologically
case where the field has angular dependence. The simplest such case
appears in the O(3) model. We consider the ansatz:
\begin{equation}
{\vec
\Phi}=(\sin\Psi_2(\rho,t)\sin\varphi,\sin\Psi_2(\rho,t)
\cos\varphi,\cos\Psi_2(\rho,t))
\end{equation}
Singlevaluedness of the field implies that $\Psi_2\rightarrow m\pi$ as
$\rho\rightarrow 0$ $m=0,\pm 1,...$
In the case of configurations with integer topological charge
($\Psi_2\rightarrow n\pi$ as $\rho\rightarrow \infty$) it may be shown
\cite{pip} that
there is an exact self-similar solution describing the evolution of
$\Psi_2$: $\Psi_2(\rho,t)=n\pi\pm\sin^{-1}(\rho/t)$ where n is any
integer.
The static energy of the above general cylindrically symmetric configuration is
\begin{equation}
E_{st}={1\over 2}\int
d^2\rho \hskip .1cm ({\vec \bigtriangledown}{\vec\Phi})^2=\pi\int d\rho\hskip
.1cm \rho \hskip .1cm (\Psi_2^{\prime 2} + {{\sin^2\Psi_2}\over
{\rho^2}})
\end{equation}
The second term in $E_{st}$ is in general logarithmically
diverging and comes from the angular gradients. Therefore, before
rescaling $\rho$ in $\Psi_2$ we must impose a cutoff in the integral.
\par
The energy of the rescaled configuration is:
$E_{st}^\alpha=\pi\int_0^{\alpha R}d\rho\hskip .1cm \rho \hskip .1cm
(\Psi_2^{\prime 2} + {{\sin^2\Psi_2}\over
{\rho^2}})$. Thus ${{\partial
E_{st}^\alpha}\over{\partial\alpha}}{\vert_{\alpha=1}}=
\pi\sin^2\Psi_2(R)+O(R^{-1})>0$
which favors decrease of $\alpha$ and therefore expansion of the
initial configuration. This is demonstrated in Fig. 2 where we show
the evolution of $\Psi_2(\rho)=0.8\pi(1-e^{-\rho})$. Similar results
were obtained with several similar initial conditions.
%\begin{figure}[htb]
%\hspace{1in}
%\scaledpicture 7.5in by 5.0in (fig2 scaled 600)
%\caption{The evolution of $\Psi_2(\rho,t_i)=0.8\pi(1-e^{-\rho})$ based on
%nonlinear O(3) model dynamics. The configuration expands in agreement
%with the analytical predictions.}
%\end{figure}
\par
The same conclusion may be reached in the O(N+1) case:\hfil\break
$
{\vec\Phi}=(\sin\Psi_N...\sin\Psi_2\sin\varphi,...,\cos\Psi_N)
$
where
$
E_{st}^\alpha=\pi\int_0^{\alpha R}d\rho\hskip .1cm \rho \hskip .1cm
(\Psi_N^{\prime 2}+\Psi_{N-1}^{\prime
2}\sin^2\Psi_N+...+{{\sin^2\Psi_2...\sin^2\Psi_N}\over{\rho^2}})
$
.This implies
$
{{\partial E_{st}^{\alpha}}\over{\partial
\alpha}}\vert_{\alpha=1}=\pi \sin^2\Psi_2(R)...\sin^2\Psi_N(R)>0
$
leading to expansion.
\par
Configurations with spherically symmetric energy density are
particularly interesting. Here we will consider global fields with
spherically symmetric energy density but with
angular dependence. By causality, fields are uncorrelated on horizon
scales and therefore fields with angular dependence are more natural
in a cosmological setup.
 The simplest such configuration
appears in the O(3) model:\hfil\break
\begin{equation}
{\vec\Phi}=
(\sin\Psi_2(r,t)\sin\lambda\theta,\sin\Psi_2(r,t)
\cos\lambda\theta,\cos\Psi_2(r,t))
\end{equation}
where $\theta$ is the polar angle and $\lambda$ is a parameter
determining the angular dependence. The static energy is
\begin{equation}
E_{st}=2\pi\int_0^Rdr\hskip .1cm r^2 (\Psi_2^{\prime 2} +
{{\lambda^2\sin^2\Psi_2}\over{r^2}})
\end{equation}
where the cutoff is imposed due
to the second diverging (in general) term. Let $E_{st}^\alpha$ be the
rescaled energy and the asymptotic behavior of $\Psi_2$ as
$r\rightarrow R$ be $\Psi_2(r)\rightarrow b\pi$. A collapsing
configuration must satisfy the following conditions:
\begin{equation}
{{\partial
E_{st}^\alpha(b)}\over{\partial\alpha}}\vert_{\alpha=1}<0
\end{equation}

\begin{equation}
{{\partial E_{st}(b)}\over{\partial b}}\leq 0
\end{equation}
The importance of (7) for collapse is obvious. Condition (8)
makes sure that the collapse will not be halted due
to the decrease of the field asymptotic value b.
By rescaling
$r\rightarrow \alpha r$ in $\Psi_2$ it is easy to show that (7) becomes:
\begin{equation}
\lambda^2(R\sin^2\Psi_2(R)-\int_0^R dr \hskip .1cm
\sin^2\Psi_2)-\int_0^R dr \hskip .1cm r^2 \hskip .1cm \Psi_2^{\prime
2}<0
\end{equation}
Condition (6) may be written:
\begin{equation}
{{\partial
E}\over{\partial b}}=R\lambda^2\sin(2b\pi)+O(1)\leq 0\Longrightarrow
{b\geq{1\over
2}}
\end{equation}
For the sake of definiteness we will consider a specific choice
for $\Psi_2$. It will be seen that the results are relatively insensitive to
this specific choice but depend crucially on the asymptotic behavior
of $\Psi_2$. Consider the linear ansatz:
$\Psi_2(r)=b\pi r$ for $r\leq 1$ and $\Psi_2(r)=b\pi$
for $r>1$. With this ansatz (7) becomes:
\begin{equation}
\lambda^2({{\sin(2b\pi)}\over{5\pi}}+\sin^2b\pi-{1\over2})-
{{(b\pi)^2}\over 3}<0
\end{equation}
The values of $b(\lambda)$ for which
${{E_{st}^\alpha(b)}\over{\partial\alpha}}\vert_{\alpha=1}=0$ (continous
line) and
${{\partial E_{st}(b)}\over{\partial b}}=0$ (dashed line) are
shown in Fig. 3.
%\begin{figure}[htb]
%\hspace{1in}
%\scaledpicture 8.51in by 5.78in (fig3 scaled 600)
%\caption{The values of $b({\lambda^2\over 2}\equiv c)$ for which
%${{E_{st}^\alpha(b)}\over{\partial\alpha}}\vert_{\alpha=1}=0$ (continous
%line) and
%${{\partial E_{st}(b)}\over{\partial b}}=0$ (dashed line).}
% \end{figure}
\par
Field configurations with b initially in region I are predicted to
immediately start collapsing since both (7) and (8) are satisfied.
Configurations in region II will
initially expand while increasing b and after they cross the line of
criticality they will start collapsing. Finally configurations in
region III do not collapse.
\par
In contrast to the topological $\pi_3$ texture, the above O(3) field
configuration is topologically trivial (i.e. can be smoothly deformed
to the vacuum) for all $\lambda$ and field asymptotic values b.
However the configuration collapses
for certain field asymptotic values. Therefore the property of field
collapse is not in general connected with topological properties.
\par
We have numerically tested the above predictions by using several
initial conditions spanning all three regions with two different
functional forms for $\Psi_2(R)$: \hfil\break
a) The linear ansatz that was used to obtain the curve
${{E_{st}^\alpha(b)}\over{\partial\alpha}}\vert_{\alpha=1}<0$ in
Fig.3\hfil\break
b)$\Psi_2(r)=b\pi(1-e^{-r})$.
\par
The results are in good agreement with
each other and they verify the analytical predictions. Here we
present a sample of the results for
the evolution with (b) as initial condition. The equation of
motion that was used for the evolution is:
\begin{equation}
{\ddot \Psi_2}={\Psi_2^{\prime\prime}}+{2\over r}\Psi_2^\prime-
{{\lambda^2 \sin {2\Psi_2}}\over {2r^2}}
\end{equation}
This equation is obtained from the full three dimensional equations of
motion using the ansatz (5).
%\begin{figure}[htb]
%\hspace{1in}
%\scaledpicture 7.5in by 5.0in (fig4 scaled 600)
%\caption{The field evolution for parameter values c=1, b=0.45}
%\end{figure}
\par
Fig. 4 shows the evolution of ansatz (b)
for parameter values $c=1$ and $b=0.45$.
As predicted the field expands while reducing the value of the
parameter $b$. The initial values of the parameters in Fig. 5 are
$c=1.0$, $b=0.57$. The collapse starts almost immediately as predicted.
Fig. 6 shows the evolution for parameter values $c=2.0$ and $b=0.54$. The
predicted evolution is initial expansion and eventual collapse once
the critical value of $b$ ($b_{cr}(c=2.0)\simeq 0.61$) is reached. This is
verified by the numerical result. We performed several similar
numerical tests for different values of the parameters $c$, $b$ and for
both ans$\ddot a$tze (a) and (b).
%\begin{figure}[htb]
%\hspace{1in}
%\scaledpicture 7.5in by 5.0in (fig5 scaled 600)
%\caption{The field evolution for parameter values c=1, b=0.57}
%\end{figure}
In Fig. 3 we summarize the results for the linear ansatz (a)
superimposed on the analytical prediction.
 Each cross (star) represents a configuration with the indicated parameter
values that
started immediately to expand (collapse). Each vector represents a
configuration that started with
parameters at the begining of the vector, initially expanded but eventually
started to collapse
when the parameters reached the values at the end of the vector.
 Clearly the analytical predictions are well verified.
The numerically obtained critical values of b
for ans$\ddot a$tze (a) and (b) were in
agreement to within about $5\%$.
%\begin{figure}[htb]
%\hspace{1in}
%\scaledpicture 8.79in by 6.04in (fig6 scaled 600)
%\caption{The field evolution for parameter values c=2, b=0.54}
%\end{figure}
\par
Even though the above results were obtained for the O(3) model it may
be shown that they apply to several other cases. For example for the
$\pi_3$ texture ansatz
\begin{equation}
{\vec\Phi}=(\sin\Psi_3\sin\theta\sin\varphi,
\sin\Psi_3\sin\theta\cos\varphi,\sin\Psi_3\cos\theta,\cos\Psi_3)
\end{equation}
we have $E_{st}=2\pi\int_0^Rdr\hskip .1cm r^2 (\Psi_2^{\prime 2} +
{{2\sin^2\Psi_2}\over{r^2}})$ and the equation of motion is (12) with
$\lambda^2=2$. Therefore the previous analysis is
applicable with $\lambda=\sqrt{2}$.
\par
Another example is the O(N+1) ansatz
$
{\vec \Phi}=(\sin \Psi_N...\sin \lambda\theta,...,\cos \Psi_N)
$
with $\Psi_N=\Psi_N(r,t)$, $\Psi_{N-1}...\Psi_2=const$. Several O(N)
configurations may be reduced to this case with appropriate
parametrization of the vacuum manifold. The static energy reduces to
(13) with $\Psi_2\rightarrow \Psi_N$. Thus the above analysis is also
applicable in this case.
\par
The fact that the property of spherical collapse is shared by many
O(N) global field configurations has important cosmological
implications. In particular, the scaling solution parameters
that determine the number of collapsing spherical knots per Hubble
volume per Hubble time are determined by the dimensionality of the
vacuum manifold. For the special case of
an $S^3$ vacuum manifold in three dimensional space
the probabilty of texture formation has been
obtained in Ref. 3 using Monte Carlo simulation and analytical
arguments (see also Ref. 9). In general this probability will vary with N.
Consequently, even though all models with broken O(N)
symmetry predict the existence of hot and cold spots in the microwave
sky, the number of predicted spots is a function of N. The
determination of this dependence could lead to a constraint in N from
purely cosmological considerations as the microwave background
distortion bounds are improving. This issue is currently under
investigation.
\par
To summarize we have shown that\hfil\break
a) Global fields with planar and cylindrical symmetry do not
collapse.\hfil\break
b) Global fields with spherical symmetry may either collapse or expand
depending on the asymptotic value of the field. \hfil\break
c) The rescaling arguments of Derrick's theorem offer in several cases,
a useful method in
determining the critical asymptotic field value $b_{crit}$.\hfil\break
We determined $b_{crit}$ in a model with broken O(3)
symmetry and showed that our analysis is applicable to several other
models. The expansion of planar and cylindrically symmetric
configurations combined with the results for spherical collapse indicates that
spherical collapse is an attractor and any distorted configuration
with the appropriate asymtpotic value will tend to restore its spherical
shape and collapse\cite{ps91}\cite{stpr91}.

\centerline{\bf Ackowledgements}
\par
I wish to thank Robert Brandenberger, Tomislav Prokopec
and Andrew Sornborger
for interesting discussions. This work was supported by a CfA Postdoctoral
Fellowship and by the U.S. DOE under grant DE-AC02-76ER03130 Tasks K $\&$ A.
\newpage
\centerline{\bf Figure Captions}
\par
{\bf Figure 1 :} The evolution of $\Psi_1(x,t_i)=0.9\pi(1-e^{-x})$ based on
 O(2) model dynamics.
\par
{\bf Figure 2 :} The evolution of $\Psi_2(\rho,t_i)=0.8\pi(1-e^{-\rho})$ based
on
nonlinear O(3) model dynamics. The configuration expands in agreement
with the analytical predictions.
\par
{\bf Figure 3 :}The values of $b({\lambda^2\over 2}\equiv c)$ for which
${{E_{st}^\alpha(b)}\over{\partial\alpha}}\vert_{\alpha=1}=0$ (continous
line) and
${{\partial E_{st}(b)}\over{\partial b}}=0$ (dashed line).
\par
{\bf Figure 4 :}The field evolution for parameter values c=1, b=0.45
\par
{\bf Figure 5 :}The field evolution for parameter values c=1, b=0.57
\par
{\bf Figure 6  :}The field evolution for parameter values c=2, b=0.54


\begin{thebibliography}{99}
\bibitem{t89} N. Turok, {\it Phys. Rev. Lett.} {\bf 63}, 2625
(1989).
\bibitem{d64} G.H.Derrick, {\it J. Math. Phys.} {\bf 5}, 1252 (1964).
\bibitem{lp91}R. Leese and T. Prokopec, 'Monte Carlo Simulation of
Texture Formation'.BROWN-HET-804 (1991).
\bibitem{p92}L. Perivolaropoulos, 'Instabilities and Interactions of
Global Topological Defects'. To appear in {\it Nucl. Phys.} B. (1992).
\bibitem{ps91}T. Prokopec, A. Sornborger, BROWN-HET-840 (1991).
\bibitem{stpr91}D.N. Spergel, N. Turok, W. H. Press, B.S. Ryden, {\it Phys.
Rev.},
{\bf D43}, (1991).
\bibitem{ts91}N. Turok and D. N. Spergel, {\it Phys. Rev. Lett.} {\bf 66}
3093 (1991).
\bibitem{bcl91} J. Borrill, E.J. Copeland, A.R. Liddle, {\it Phys. Lett.},
{\bf B258}, 310 (1991).
\bibitem{pip}L. Perivolaropoulos, unpublished.
\end{thebibliography}
\end{document}